\documentclass{ws-ijmpd}
\usepackage[super,compress]{cite}
\usepackage{graphicx}
\usepackage{rotating}
\usepackage{lscape}
\begin{document}

\markboth{VINCENZO GALLUZZI, MARCELLA MASSARDI}
{THE POLARIMETRIC MULTI-FREQUENCY RADIO SOURCES PROPERTIES}
%%%%%%%%%%%%%%%%%%%%% Publisher's Area please ignore %%%%%%%%%%%%%%%
%
\catchline{}{}{}{}{}
%
%%%%%%%%%%%%%%%%%%%%%%%%%%%%%%%%%%%%%%%%%%%%%%%%%%%%%%%%%%%%%%%%%%%%

\title{THE POLARIMETRIC MULTI-FREQUENCY RADIO SOURCES PROPERTIES}

\author{VINCENZO GALLUZZI}
\address{INAF, Osservatorio di Radioastronomia, Via Gobetti 101\\
Bologna, 40129, Italy\\
vgalluzzi@ira.inaf.it}
\address{Dipartimento di Fisica e Astronomia, Universit\`a di Bologna, via Ranzani 1\\
Bologna, 40126, Italy\\
vincenzo.galluzzi@unibo.it}

\author{MARCELLA MASSARDI}

\address{INAF, Osservatorio di Radioastronomia, Via Gobetti 101\\
Bologna, 40129, Italy\\
massardi@ira.inaf.it}

\maketitle

\begin{history}
\received{Day Month Year}
\revised{Day Month Year}
\end{history}

\begin{abstract}
The polarization properties of extragalactic radio sources at frequencies higher than $20\,$GHz are still poorly constrained. However, their characterization would provide invaluable information about the physics of the emission processes and is crucial to estimate their contamination as foregrounds of the polarized cosmic microwave background (CMB) angular power spectrum on scales $\lesssim 30\,$arcmin. In this contribution, after summarizing the state-of-the-art of polarime\-tric observations in the millimetric wavelength bands, we present our observations of a complete sample of 53 sources with $S_{20\,{\rm GHz}}>200\,$mJy carried out with the Australia Telescope Compact Array between $5.5$ and $38\,$GHz. The analysis clearly shows that polarization properties cannot be simply inferred from total intensity ones, as the spectral behaviors of the two signals are typically different.
\end{abstract}


\ccode{PACS numbers:}

%\tableofcontents

\section{The state-of-the-art in polarimetric observations}	

Bright extragalactic radio source samples are mostly associated to the nuclei of active galaxies. A widely held view is that their spectra result from a superposition of a number of compact regions of different sizes, self-absorbed at different frequencies (e.g. see Ref.~\refcite{Marscher1980} and Ref.~\refcite{DeZotti2010} for a review), associated to knot-like Doppler-boosted structures along a relativistic jet, powered by the central nuclear activity.\cite{Blandford1979}

The characterization of the total intensity emission for large samples of radio source population is only a recent achievement, and still open to discussion. Wide-area surveys are necessary to achieve statistics on the bright less numerous samples, while high sensitivity is needed to explore the faintest samples. Only the former were available in the catalogs extracted from full-sky maps from satellite missions like WMAP (with a completeness limit of $\simeq 1\,$Jy at 23 GHz\cite{Argueso2003,DeZotti2005,Wright2009,Gold2011}) and Planck (with a completeness limit of $\simeq 500\,$ mJy at $30\,$GHz \cite{PlanckCollaboration2015}). New broad-band correlators made high sensitivity available for interferometric observations up to the mm and sub-mm regimes, allowing wide-area deep surveys to be carried out also from the ground at frequencies above $\sim 10\,$GHz. Thus, the Australia Telescope 20 GHz (AT20G) Survey \cite{Murphy2010,Massardi2011a} covered the full Southern sky with 91\% completeness above $100\,$mJy and $79\%$ completeness above $50\,$mJy in regions south of declination $-15^\circ$. 

By combining the ground- and satellite-based instrumental capabilities, several authors \cite{Massardi2008,Sajina2011,Chen2013,Massardi2015} recently provided a broad-band view of the total intensity emission of the bright radio source population: it seems to be dominated by relatively young compact objects. A double power-law model is adequate to describe spectral behaviors for more than $2$ decades in frequency: this indicates that a single dominant component is responsible for the (typically optically thin) emission above $\sim 30\,$GHz. Emission is typically variable: objects brighter than $500\,$mJy (at $20\,$GHz) on average vary their flux density (at $20\,$GHz) more than $6\%$ over a 6 months lag, and the rate grows with frequency.

The study of the polarized emission both in frequency and space would help describing the dynamics of the jets. Synchrotron emission of each component is intrinsically highly linearly polarized\cite{Ginzburg1969} (up to $70-80\%$) but typically observed integrated polarized fractions for compact extragalactic radio sources are rarely as high as $\sim 10$\%. They are, in fact, the result of vector averaging along the line of sight (depolarization is mostly induced by differential Faraday rotation)\cite{Burn1966}, within the observing resolution element (in this case unresolved magnetic sub-structures cause depolarization), and within the band of the polarized components emission. Hence, polarimetric observations typically still constitute an observational challenge because of the requested high sensitivity, calibration accuracy, and the detailed knowledge of instrumental properties and systematics (see Massardi {\it et al.} in this volume).
Table 1 lists some of the available multi-frequency compilations, surveys and complete samples follow-up that include polarimetric information: it does not claim to be an exhaustive picture and aims to trace the basic references in the following discussion.

For the above mentioned reasons, our current knowledge of polarimetric properties of radio sources mostly rely on $<10$ GHz selected samples including the NVSS survey\cite{Condon1998} that covered the full sky above $-45^\circ$, remaining complete down to $2.5\,$mJy in total intensity and with $\sigma_P\sim 0.2\,$mJy in polarized emission. 

\begin{landscape}
\begin{table}
   %\centering
   \tbl{Summary of some of the surveys in polarization available at radio frequencies (update to Table 3 in Tucci {\it et al.}\cite{Tucci2004}, 2004).}
{
   \begin{tabular}{lccl}
\hline
\textbf{References}            &\textbf{Frequency (GHz)}&\textbf{\# sources}& \textbf{Notes}     \\
\hline
Eichendorf \& Reinhardt (1979)\cite{Eichendorf1979}                 & $[0.4,\,15]$                    & $510$  & compilation of multi--frequency data\\
Tabara \& Inoue (1980)\cite{Tabara1980}                                 & $[0.4,\,10.7]$                  & $1510$ & compilation of multi--frequency data\\
Simard-Normandin {\it et al.} (1981)\cite{Simard-Normandin1981a,Simard-Normandin1981b} & $[1.6,\,10.5]$  & $555$  & compilation of multi--frequency data\\
Perley (1982)\cite{Perley1982}                                                 & $1.5,\,4.9$                     & $404$  & compilation of multi--frequency data\\
Rudnick {\it et al.} (1985)\cite{Rudnick1985}                                 & $[1.4,\,90]$                           & $20$   & compilation of multi--frequency data\\
Aller {\it et al.} (1992)\cite{Aller1992}                                 & $4.8,\,8.0,\,14.5$                    & $62$   & $90\%$ complete sample with $S_{5{\,\rm GHz}}>1.3\,$Jy\\
Okudaira {\it et al.} (1993)\cite{Okudaira1993}                                & $10$                           & $99$   & flat-spectrum sources with $S_{5{\,\rm GHz}}>0.8\,$Jy\\
Nartallo {\it et al.} (1998)\cite{Nartallo1998}                                & $273$                           & $26$   & compilation of flat-spectrum radio sources\\
Condon {\it et al.} (1998) - NVSS\cite{Condon1998}                        & $1.4$                           & $\sim 2\times10^{6}$ & $100\%$ complete survey down to $S_{1.4{\,\rm GHz}}>2.5\,$mJy\\
Aller {\it et al.} (1999)\cite{Aller1999}                                  & $4.8,\,8.0,\,14.5$                    & $41$   & BLLac sources\\
Fanti {\it et al.} (2001)\cite{Fanti2001}                           & $4.9, 8.5$           & $87$ & CSS sample with $S_{0.4{\,\rm GHz}}>0.8\,$Jy\\
Lister (2001)\cite{Lister2001}                                 & $43$                            & $32$   & $90\%$ complete sample with $S_{5{\,\rm GHz}}>1.3\,$Jy\\
Klein et al (2003)\cite{Klein2003}                            & $1.4,\,2.7,\,4.8,\,10.5$         & $192$  & compilation of detections of the B3-VLA survey\\
Ricci {\it et al.} (2004)\cite{Ricci2004}                                  & $18.5$                                 & $250$  & complete sample with $S_{5\,\rm GHz} > 1\,$Jy \\
Jackson {\it et al.} (2007)\cite{Jackson2007}                                 & $8.4$                                         & $\sim 16000$ & JVAS-CLASS surveys\\
Massardi {\it et al.} (2008) AT20G-BSS\cite{Massardi2008}         & $4.8,\,8.6,\,20$                        & $320$  & AT20G bright sample\\
Lopez-Caniego {\it et al.} (2009)\cite{Lopez-Caniego2009}                        & $23,\,33,\,41$                                & $22$ & polarization detections in WMAP maps\\
Jackson {\it et al.} (2010)\cite{Jackson2010} & $8.4,\,22,\,43$         & $230$ & WMAP sources follow-up\\
Murphy {\it et al.} (2010) AT20G\cite{Murphy2010} & $4.8,\,8.6,\,20$ & $5890$ & $93\%$ complete survey with $S_{20{\,\rm GHz}}>40\,$mJy\\
%Agudo {\it et al.} (2010)\cite{Agudo2010}                                        & 86 & 146  & complete sample of flat-radio-spectrum with $S_{86{\,GHz}}>1\,$Jy \\
Trippe {\it et al.} (2010)\cite{Trippe2010} & $[80,\,267]$ & $86$ & complete sample with $S_{90\,\rm GHz}> 0.2\,$Jy \\
Battye {\it et al.} (2011)\cite{Battye2011} & $8.4,\,22,\,43$         & $230$ & WMAP sources follow-up\\
Sajina {\it et al.} (2011)\cite{Sajina2011}                                & $4.8,\,8.4,\,22,\,43$                & $159$ & AT20G sources follow-up\\
Massardi {\it et al.} (2013)\cite{Massardi2013}                                                 & $4.8,\,8.6,\,18$                         & $193$ & complete sample with $S_{20{\,\rm GHz}}>500\,$mJy\\
Agudo {\it et al.} (2014)\cite{Agudo2014}                                        & $86,\,229$ & $211$ & complete sample of flat-spectrum sources with $S_{86{\,\rm GHz}}>1\,$Jy \\
Planck Collaboration (2015)\cite{PlanckCollaboration2015} & $30,\,44,\,70$ & $122,\,30,\,34$ & polarization detections in Planck LFI maps (PCCS2)\\ 
                            & $100,\,143,\,217,\,353$ & $20,\,25,\,11,\,1$ & polarization detections in Planck HFI maps (PCCS2)\\ 
\hline
   \end{tabular} 
\label{Tab1}}
%\caption{Summary of some of the surveys in polarization available at radio frequencies (update to Table 3 in Tucci {\it et al.}\cite{Tucci2004} (2004).}
\end{table}
\end{landscape}

Noticeable exceptions are few high frequency-selected samples that are typically limited to flat-spectrum or bright objects.

The polarization of WMAP sources has been investigated by L{\'o}pez-Caniego \textit{{\it et al.}} (2009, Ref.~\refcite{Lopez-Caniego2009}) by using WMAP data: 14 extragalactic objects were significantly detected in polarization. Slightly larger samples were detected in the Planck maps and recorded in the ``Planck Catalogue of Compact Sources''\cite{Planck Collaboration 2015} (PCCS, 2nd version), listing 122 detections down to a minimum polarized flux density of $117\,$mJy at $30\,$GHz but complete only to $0.6\,$Jy.
Ground-based follow-up observations of a complete sample of $203$ WMAP sources were carried out with the VLA by Jackson {\it {\it et al.}} (2010, Ref. ~\refcite{Jackson2010}) and Battye {\it {\it et al.}} (2011, Ref.~\refcite{Battye2011}): polarized emission was detected for $123$, $169$ and $167$ objects at $8.4$, $22$ and $43\,$GHz, respectively.

The Plateau de Bure Interferometer (PdBI) observations of Trippe {\it {\it et al.}} (2010, Ref.~\refcite{Trippe2010}) of a $S_{90\, \rm GHz}>200$\,mJy complete sample of 86 sources found an average fractional polarization level of $\simeq 2-7\,\%$, higher for BLLac ($\simeq 7\,\%$) than for QSO ($\simeq 5\,\%$) or Seyfert galaxies ($\simeq 3\,\%$). The size scales relevant for the polarization emission measurements are found to be comparable to those of interest for total intensity flux density measurements.

The full AT20G catalog\cite{Murphy2010,Massardi2011a} includes the $20\,$GHz polarized intensity for $768$ sources, $467$ of which also have simultaneous polarization detections at $5$ and/or $8\,$GHz, out of a total of $5890$ sources. The detection limit is defined as $\max(3\,\sigma,0.01S_{20\, \rm GHz},6\,\hbox{mJy})$. Sadler {\it et al.} (2006, Ref.~\refcite{Sadler2006}) presented polarization measurements for a sample of $173$ AT20G sources brighter than $S_{20\, \rm GHz}=100\,$mJy: $129$ ($\simeq 75\%$) were detected at $20\,$GHz, with a median fractional polarization of $2.3\%$. Massardi {\it et al.} (2008, Ref.~\refcite{Massardi2008}) discussed the polarization properties of the AT20G bright sample ($S_{20\, \rm GHz}\ge 500\,$mJy), finding $213$ polarization detections  ($\ge 3\,\sigma$) at 20 GHz out of a total of 320 sources ($\simeq 67\%$), with a median fractional polarization of $2.5\%$ at 20 GHz (confirmed also by Refs.~\refcite{Jackson2010,Battye2011}). The spectral indices in total intensity and in polarization were found to be similar on average, but there were several objects for which the spectral shape of the polarized emission is substantially different from the spectral shape in total intensity. Several studies of radio source polarization, mostly for samples selected at 1.4 GHz and dominated by steep-spectrum objects, have reported indications that the polarization degree increases with decreasing flux density \cite{Mesa2002,Tucci2004,Taylor2007,Grant2010,Subrahmanyan2010}. Massardi {\it et al.} (2013, Ref.~\refcite{Massardi2013}) analyzed high-sensitivity polarization observations (in the $4.8-20\,$GHz frequency range) of a complete AT20G bright ($S_{20\, \rm GHz}>500\,$mJy) source sub-sample and found no statistically significant relationship between the polarization fraction and the total intensity flux density, and no clear indication of trends of fractional polarization with frequency, up to $20\,$GHz. 

Sajina {\it et al.} (2011, Ref.~\refcite{Sajina2011}) obtained polarization measurements with the VLA at $4.86$, $8.46$, $22.46$, and $43.34\,$GHz of $159$ out of the $\sim 200$ AT20G radio galaxies with $S_{20\, \rm GHz}\ge 40\,$mJy in an equatorial field of the Atacama Cosmology Telescope survey: polarized flux was detected at $> 95\%$ confidence level for $141$, $146$, $89$, and $59$ sources, from low to high frequencies. The measured polarization fractions are typically $<5\%$, although in some cases they are measured to be up to $\simeq 20\%$. They find indications of increasing polarization fraction with frequency (confirmed also by Ref.~\refcite{Agudo2010} in the $15-90\,$GHz range and by Ref.~\refcite{Agudo2014} in the $86-229\,$GHz range). This trend is stronger for steeper spectrum sources as well as for the lower flux density sources. 

Finally, it has been argued that the ordering of magnetic fields should increase in the inner regions, giving a higher polarization degree at higher frequency\cite{Tucci2004}, while Faraday depolarization should affect more the lower frequency observations. 

However, it seems clear that extrapolations from low frequencies ($<20\,$GHz) or from total intensity emission are inadequate to model the radio source contribution in CMB polarization maps (see Ref.~\refcite{QUIET2014}) and that future surveys will benefit CMB polarization experiments by statistically characterizing radio source populations down to lower flux density limits. On the one hand, in fact, foreground sources are expected to be the most relevant contaminant of the CMB angular power spectrum for angular scales up to $\sim 30\,$arcmin in the $70-100\,$GHz frequency range, where Galactic foregrounds (Galactic synchrotron and thermal dust emissions) are, instead, at a minimum. On the other hand, radio mm-band facilities have reached so far sensitivities that, in principle, are useful to constraining models that predict cosmic polarization rotation (CPR), as a result of violations of fundamental principles (such as the Einstein Equivalence Principle, the Lorentz Invariance or the CPT symmetry)\cite{NiWT2008,Kostelecky2009,diSeregoAlighieri2015}, at the same level or even better than constraints coming from CMB observations.\cite{POLARBEAR2015}. Furthermore, multi-steradian samples of high frequency-selected polarized sources are also important for identifying suitable calibrators for CMB polarimetric experiments and upcoming millimeter-wave telescopes. See Massardi {\it et al.} in this volume for a short review about the role of radio sources in polarimetric cosmological investigations. 

Broad frequency range and multi-epoch observations of low flux density-limited samples are now needed to complement the view of radio source population pro\-perties in polarization and provide samples useful for cosmological studies, like those that are being carried out in the framework of the Planck-ATCA Co-eval Observations project\cite{Massardi2015} for a $S_{20\, \rm GHz}>200\,$mJy sample selected in the AT20G catalog. The project is briefly summarized in the following section.

\section{The Planck-ATCA Co-eval Observations project}
The Planck-ATCA Co-eval Observations (PACO, Refs. in \refcite{Massardi2011b,Bonavera2011,Bonaldi2013,Massardi2015}) project, carried out with the Australia Telescope Compact Array (ATCA), observed 464 objects drawn from the Australia Telescope survey at $20\,$GHz (AT20G, Refs. in \refcite{Murphy2010,Massardi2011a}) between $4.5$ and $40\,$GHz (below and overlapping with the two lower {\it Planck} frequency bands) in $65$ epochs during the period July 2009--August 2010, corresponding to the first two sky surveys of the satellite nominal mission. Source were scheduled for ATCA and Planck observations to occur within $10$ days of each other in order to minimize source variability effects.
The sample is made of $3$ partially overlapping sub-samples, namely the {\it bright sample} ($S_{20\,\rm GHz}>500\,$mJy), the {\it faint sample} ($S_{20\,\rm GHz}>200\,$mJy) and the {\it spectrally-selected sample}, plus one of potentially strongly variable objects identified among the ATCA calibrators database. 

The main goal of the project was to characterize the radio source population observed by the Planck satellite down to well below its detection limits (down to a factor of $5$) in a broad frequency range and on different epochs. Thanks to the combination of ATCA and Planck detected flux densities, we reconstructed the spectral properties of the sources in the $5-217\,$GHz frequency range. 
The vast majority ($\simeq 91\,$\%) of the sources show a remarkably smooth spectrum, well described by a double power law over the whole range. This translates in several compact objects showing a peaked spectrum with an optically thin steep behavior for frequencies $\gtrsim 30\,$GHz, probably associated to recently emitted components. This turns at odds with the criterion, well established at cm-wavelengths, according to which flat spectra are usually associated to compact sources while steep emitted synchrotron is indicative of extended sources. No further synchrotron break is observed at high frequencies. Assuming a continuous injection model\cite{Kardashev1962,Murgia2002}, our finding can be considered an indication that the sources are young, i.e. $\tau_{\rm syn}$ no more than $10^{3-4}\,$yr.\cite{Massardi2015}     

\subsection{Polarimetric follow-up}
In September 2014 a high sensitivity ($\sigma_P\simeq 0.2\,$mJy) polarimetric observation in $3$ correlator spectral setups (that allow $2\times2\,$GHz simultaneous bands centered at $5.5-9, 18-24$ and $33-38\,$GHz) was performed with ATCA (Massardi, project C2922) for a complete sub-sample of $53$ extragalactic compact radio sources of the faint ($S_{20\,\rm GHz}>200\,$mJy) PACO sample\cite{Bonavera2011}, covering the Southern Ecliptic Pole region (ecliptic latitude $<-75^\circ$). 

The choice of the region of the sky was motivated in PACO project by the Planck scanning strategy, which gives maximum sensitivity near the ecliptic poles. 

The array configuration was the H214, an hybrid with a nominal spatial resolution ($\simeq \lambda/b_{max}$) that spans from $8$ to $55\,$arcsec for the observed frequencies.\footnote{Indeed, taking into account the longest baseline with the sixth antenna ($\simeq 4.5\,$Km) the spatial resolution spans from $0.4$ to $2.5\,$arcsec but in this case the achievable sensitivity is reduced.} All the objects in the sample are point-like.

The project was allocated in $12\,$h ({\it i.e.} $\simeq 4\,$h per band, including overheads and calibration), during which weather conditions were very good. Every object has been observed for each spectral setup at least for $2\times 1.5\,$min cuts, that is estimated to be enough to reach the theoretical noise level of $0.2\,$mJy in polarized emission.

\begin{figure}[pb]
\begin{minipage}{0.49\textwidth}
\centerline{\psfig{file=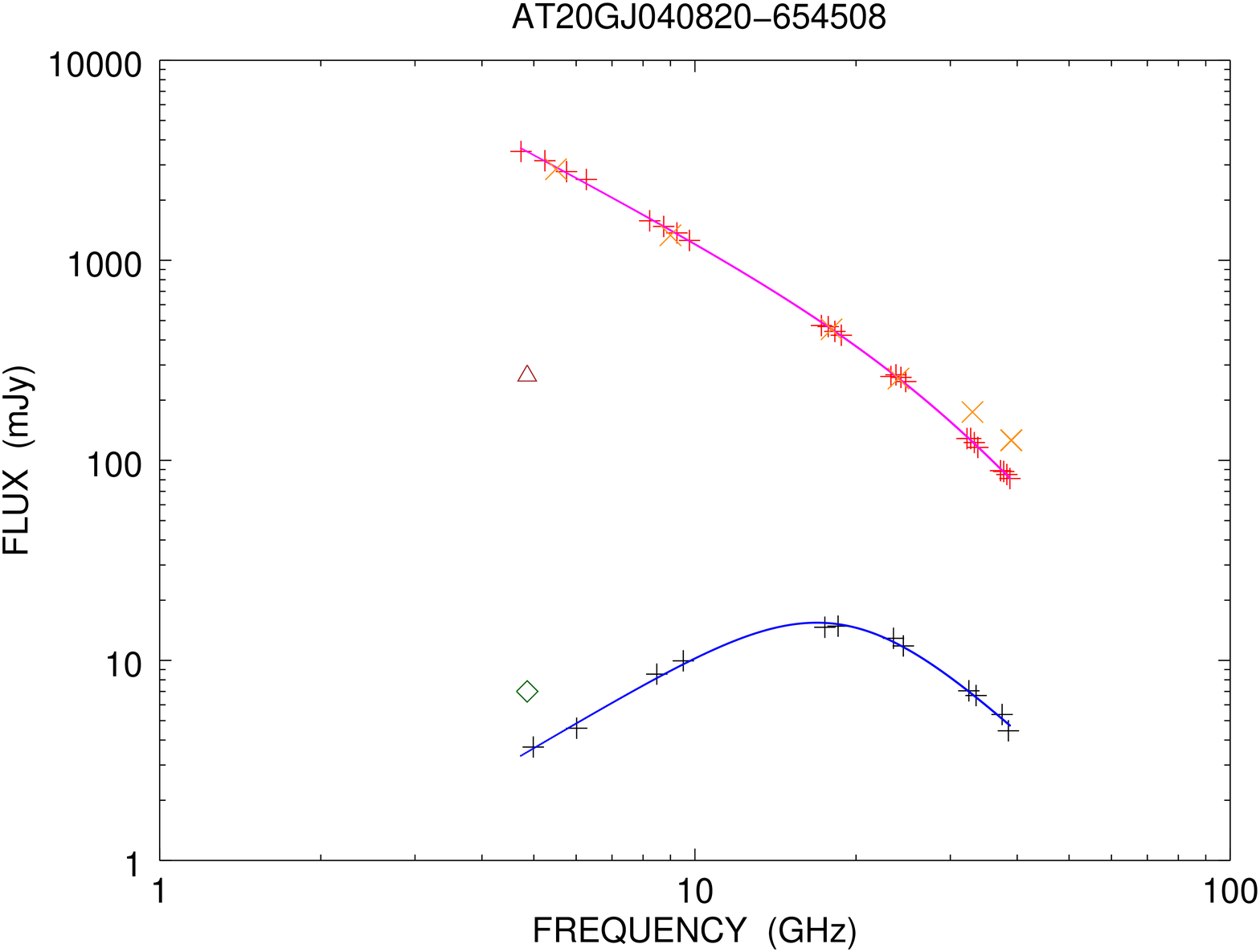,width=1.0\textwidth}}
\end{minipage}
\begin{minipage}{0.49\textwidth}
\centerline{\psfig{file=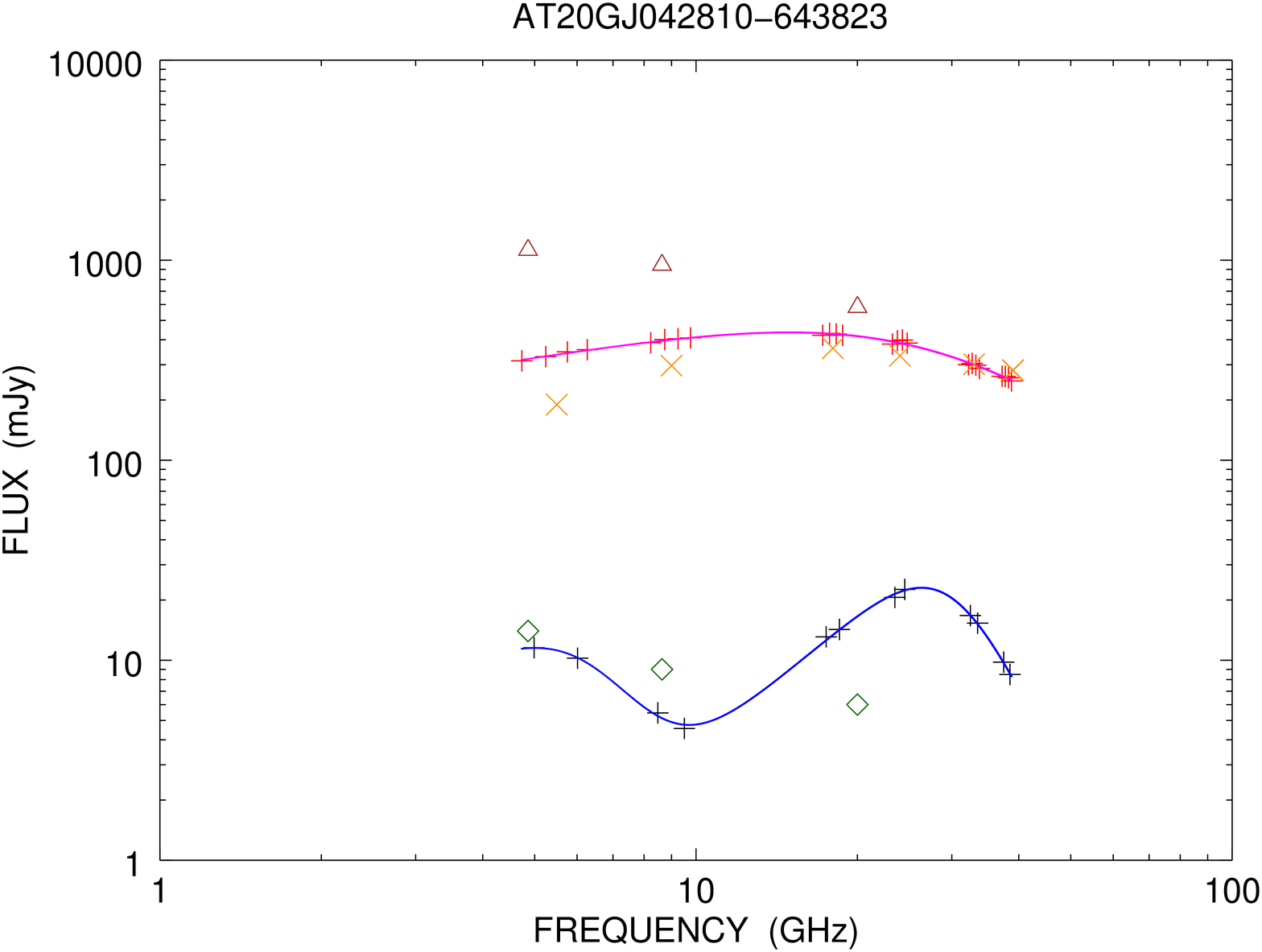,width=1.0\textwidth}}
\end{minipage}
\begin{minipage}{0.49\textwidth}
\centerline{\psfig{file=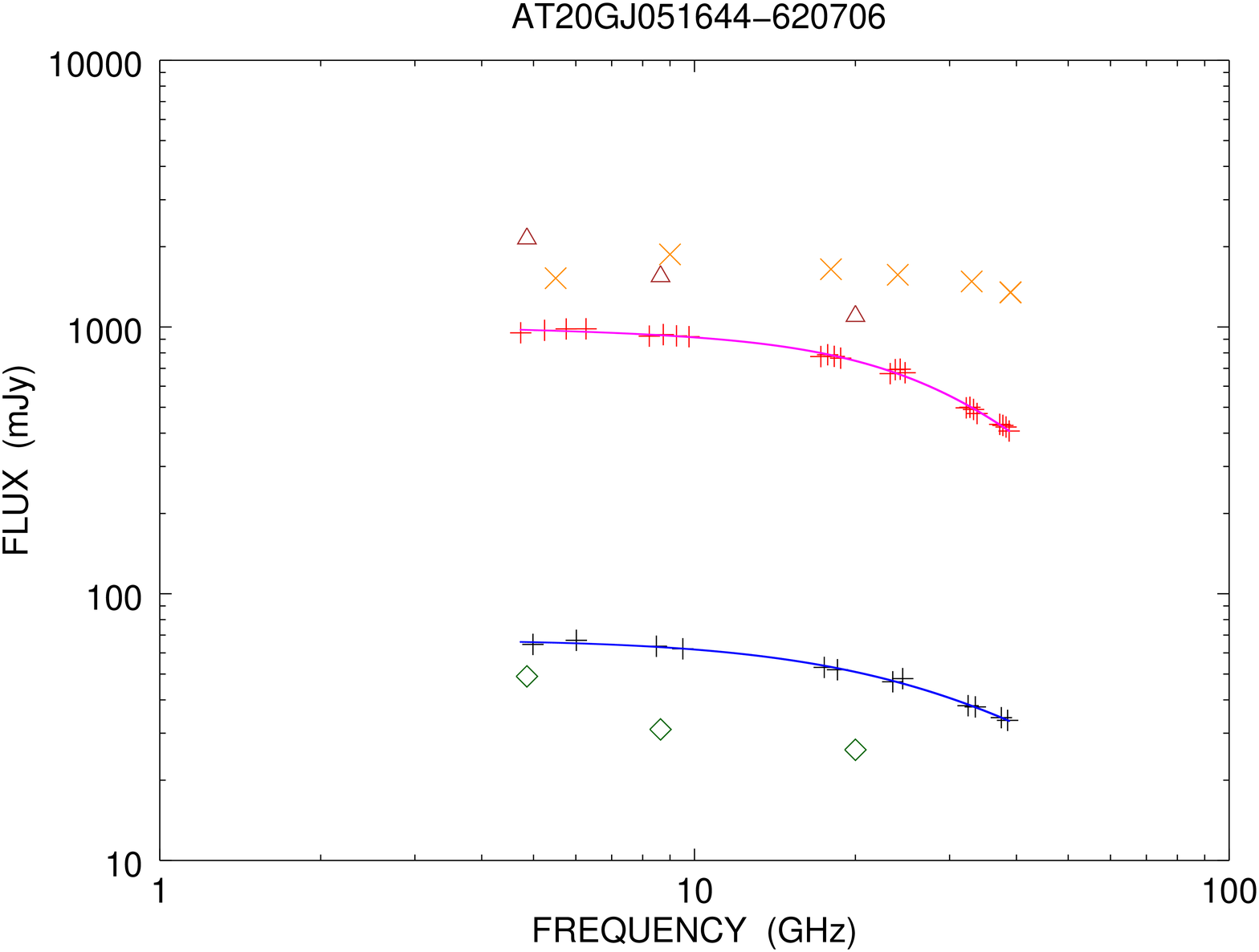,width=1.0\textwidth}}
\end{minipage}
\begin{minipage}{0.49\textwidth}
\centerline{\psfig{file=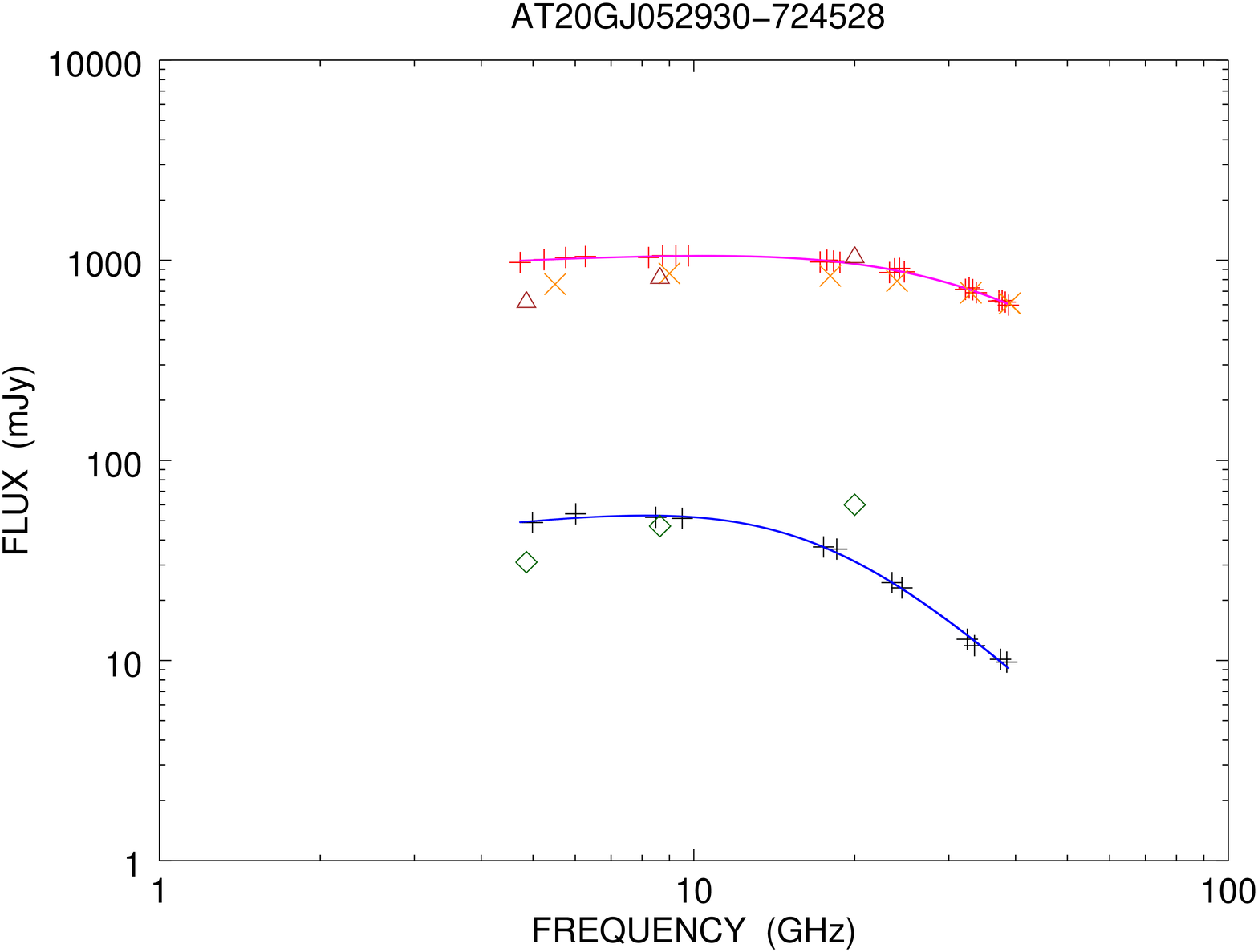,width=1.0\textwidth}}
\end{minipage}
\begin{minipage}{0.49\textwidth}
\centerline{\psfig{file=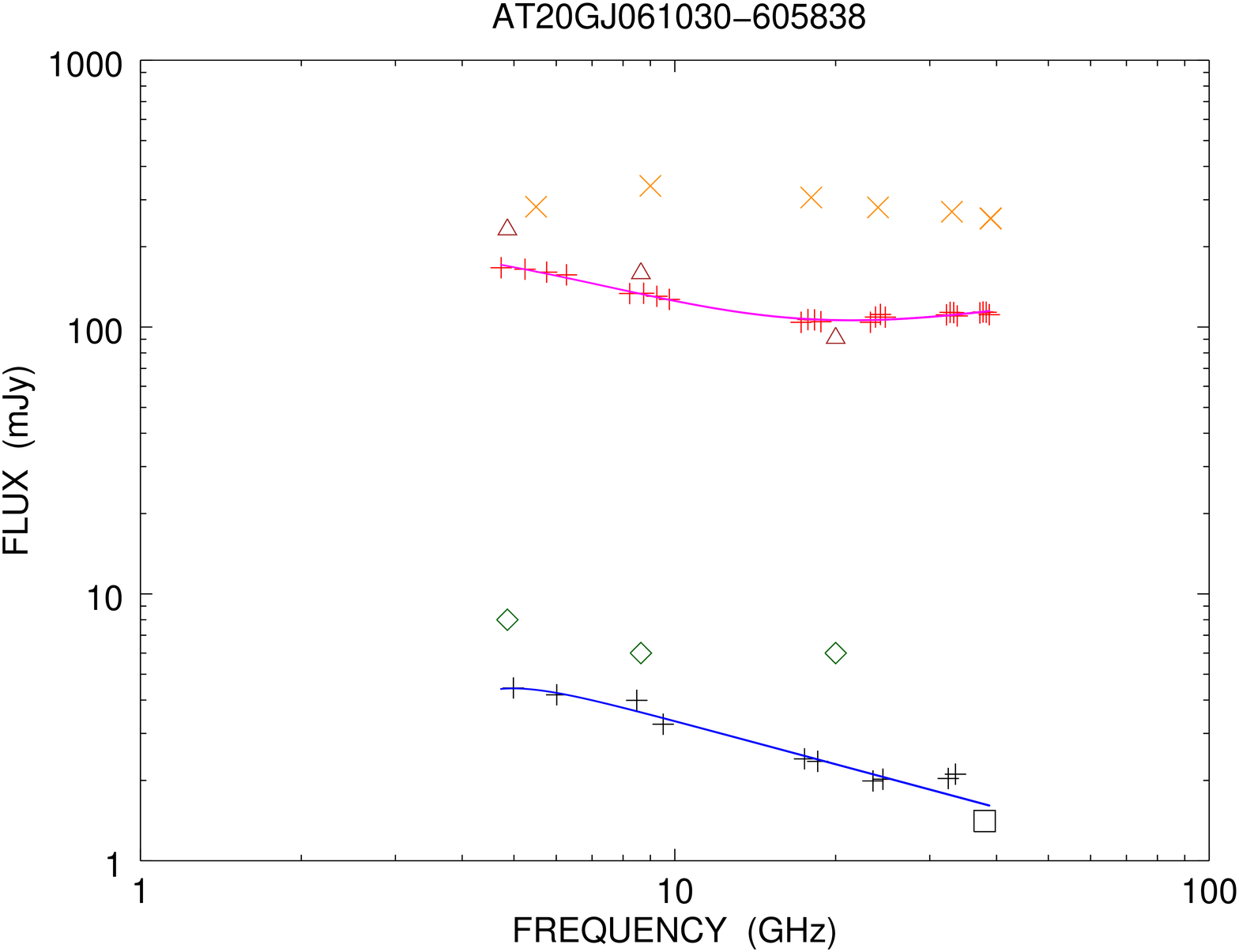,width=1.0\textwidth}}
\end{minipage}
\begin{minipage}{0.49\textwidth}
\centerline{\psfig{file=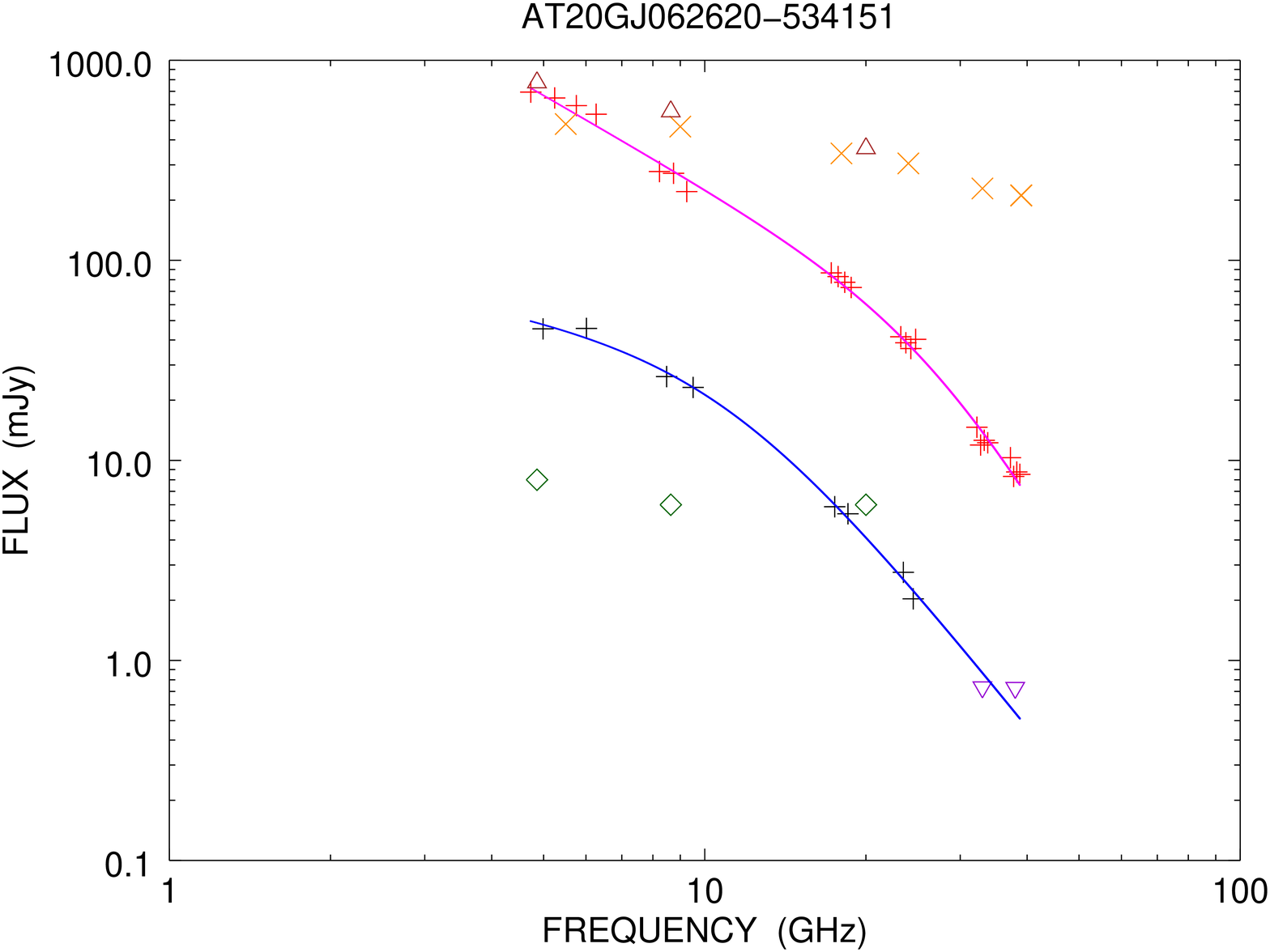,width=1.0\textwidth}}
\end{minipage}
\caption{Spectra in total intensity and polarization (error bars are not displayed since they are smaller than plotting symbols) for some sources drawn from the $53$ of the faint PACO sample, observed in September 2014. {\bf Total intensity:} red crosses indicate total intensity ATCA September 2014 observations (each point represents a $512\,$MHz-bin) and magenta solid lines indicate fitting curves. Median PACO flux densities (July 2009--August 2010) are indicated with orange ``x''. AT20G observations (best epoch in 2004--2008) are reported with brown triangles. {\bf Polarization:} black crosses and black squares indicate September 2014 observations (each point represents a $1\,$GHz-bin and the $2\,$GHz-full bandwidth, respectively) and blue solid lines indicate fitting curves. AT20G observations (best epoch in 2004--2008) are reported with green diamonds. Upper limits are shown as violet downwards triangles.\label{f1}}
\end{figure}

Data reduction was performed with MIRIAD software following standard polarimetric calibration procedures.\footnote{According to ATCA Users' Guide: www.narrabri.atnf.csiro.au/observing/users\_guide.} 
The outcome are high sensitivity spectral profiles in total intensity and polarization (some of them are reported in Fig. \ref{f1}, together with past PACO and AT20G observations available). 
The detection rate in polarization is about $90\%$ at $5\,\sigma$.

\begin{figure}[pb]
\centerline{\psfig{file=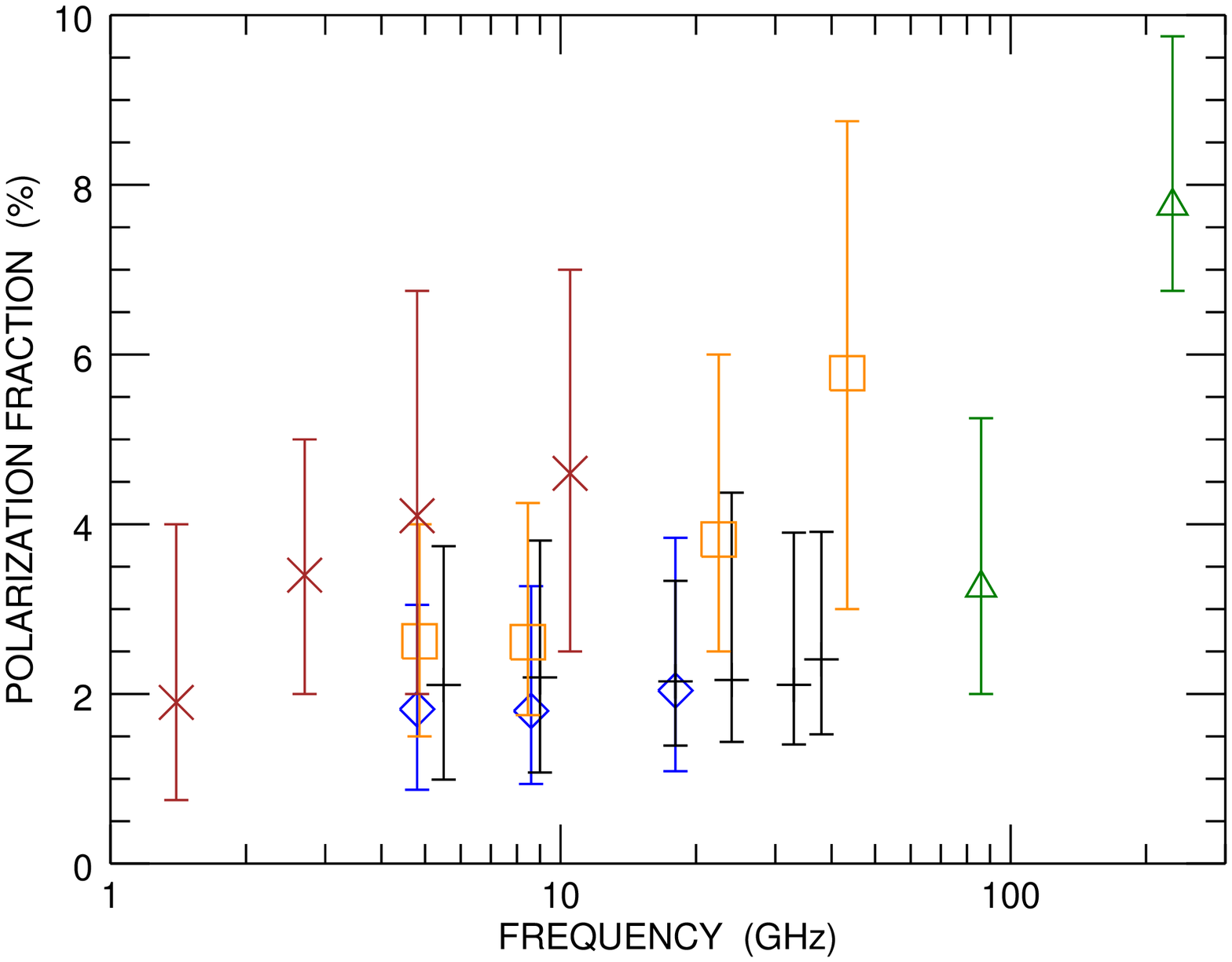,width=1.0\textwidth}}
\caption{Median fractional polarizations (error bars report inter-quartile distances)  at different frequencies for our observations (black crosses) and some other samples (reported also in the table \ref{Tab1}): the AT20G survey\cite{Murphy2010} (blue diamonds), Klein {\it et al.}\cite{Klein2003} (red ``x''), Agudo {\it et al.}\cite{Agudo2014} (green triangles), and Sajina {\it et al.}\cite{Sajina2011} (orange squares).\label{f2}}
\end{figure}

Similar to what was found in past analysis, about $96\%$ of the objects total intensity behavior can be fitted with a double power law. 
Polarimetric spectra, instead, typi\-cally show a different and sometimes more complex behavior with respect to total intensity (see Fig. \ref{f1}), in agreement with Massardi {\it {\it et al.}} findings (2013, Ref.~\refcite{Massardi2013}).
As found in total intensity, a steepening of spectra is observed at higher frequencies both in total intensity and in polarization. The median fractional polarization is $\simeq 2.1\%$ at $18\,$GHz  (in agreement with Refs.~\refcite{Sadler2006,Massardi2008,Battye2011}). A tiny trend with frequency is observed for the whole sample (see Fig.~\ref{f2}).
A comprehensive analysis of the data, including spectral classifications and source counts in polarization will be presented in a future paper (Galluzzi {\it {\it et al.}}, in preparation). 

The project will be complemented by a second observational session scheduled with ATCA in March 2016 to observe 106 sources in the faint PACO sample with ecliptic latitude $<-65^\circ$ (i.e. including the $53$ already observed in the previous epoch) at all the already observed frequencies and to extend down to $\simeq 2\,$GHz, and by an ALMA Cycle-3 project, accepted with high-ranking, to re-observe a complete sub-sample of $31$ objects at $\simeq 100\,$ GHz, that will provide interesting hints for cosmological applications, as described in Massardi {\it et al.} (this volume).

\section{Cosmological outreach}
As already stressed in the first section, extrapolations from low frequencies ($<20$ GHz) are inadequate to model the radio source contribution in CMB polarization maps \cite{QUIET2014} and future surveys will bene\-fit CMB polarization experiments by statistically characterizing radio source populations, as those proposed in our ALMA-Cycle 3 proposal.
Moreover those stable, brighter and highly polarized sources of our sample are potentially of interest as calibrators for many ground-based facilities looking for primordial B-modes in the CMB angular power spectrum or cosmic polarization rotation, for which a clear detection has not been claimed so far and available upper limits are about $1^\circ$\cite{diSeregoAlighieri2015}. Indeed, recently the POLARBEAR Collabo\-ration reports for the first time a sub-degree upper limit (see Ref. \refcite{POLARBEAR2015}). From our data we can assess a total error on polarization angle of $\simeq 3^\circ$, by assuming that the calibration error for polarized fluxes (estimated $\simeq 10\%$) is equally distributed among Stokes parameter $Q$ and $U$. However, recent measurements of polarization angles performed with VLA and ATCA are found to be in agreement within $\pm 2 ^\circ$\cite{Partridge2015}, which is a current conservative estimate of the absolute calibration error on this quantity.

\section{Acknowledgments}
We acknowledge financial support by the Italian {\it Ministero dell'Istruzione, Universit\`a e Ricerca} through the grant {\it Progetti Premiali 2012--iALMA} (CUP C52I13000140001). We thank the anonymous referee for the useful comments. We thank the staff at the Australia Telescope Compact Array site, Narrabri (NSW), for the valuable support they provide in running the telescope and in data reduction. The Australia Telescope Compact Array is part of the Australia Telescope which is funded by Commonwealth of Australia for operation as a National Facility managed by CSIRO.
%\begin{thebibliography}{000} %for 3 digits
%\begin{thebibliography}{00}  %for 2 digits

\end{document}